\documentclass[11pt]{article}

\usepackage{graphicx}
\usepackage{amssymb}
\usepackage{amsmath}
\usepackage{upgreek}
\usepackage{multirow}
\usepackage[latin1]{inputenc}
\usepackage[labelfont=bf,labelsep=period,justification=raggedright]{caption}
\usepackage{color}
\usepackage[
 breaklinks=true
 ,colorlinks=true
 ,linkcolor=blue
 ,citecolor=blue
 ,backref=none
 ,pagebackref=false
 ]{hyperref} 
\usepackage{blindtext}
\usepackage[sc]{mathpazo}	
\usepackage{cite}
\usepackage{authblk}
\usepackage{setspace} 
\usepackage{longtable}
\usepackage{fancyhdr}
\usepackage{placeins}
\makeatletter
\renewcommand{\@biblabel}[1]{\quad#1.}
\renewcommand\subsection{\@startsection{subsection}{2}{0mm} {-\baselineskip}{0.5\baselineskip}{ \bf \itshape}} 
\makeatother

\date{\today}

\begin{document}

\title{Aspiration Dynamics of Multi-player Games in Finite Populations}

\author[1]{Jinming Du \thanks{jmdu@pku.edu.cn}}
\author[2]{Bin Wu \thanks{bin.wu@evolbio.mpg.de}}
\author[3,4,5]{Philipp M.~Altrock \thanks{paltrock@jimmy.harvard.edu}}
\author[1]{Long Wang \thanks{longwang@pku.edu.cn}}

\affil[1]{{\small Center for Systems and Control, College of Engineering, Peking University, Beijing 100871, People's Republic of China}}
\affil[2]{{\small Research Group for Evolutionary Theory, Max Planck-Institute for Evolutionary Biology, August-Thienemann-Stra{\ss}e 2, 24306 Pl\"{o}n, Germany}}
\affil[3]{{\small Department of Biostatistics and Computational Biology, Dana-Farber Cancer Institute, 450 Brookline Ave, CLSB 11007, Boston, MA 02215, USA}}
\affil[4]{{\small Department of Biostatistics, Harvard School of Public Health, Boston, MA 02215, USA}}
\affil[5]{{\small Program for Evolutionary Dynamics, Harvard University, Cambridge, MA 02138, USA}}

\date{\today}

\maketitle

\newpage
\begin{abstract}
Studying strategy update rules in the framework of evolutionary game theory,
  one can differentiate between imitation processes and aspiration-driven dynamics.
In the former case,
  individuals imitate the strategy of a more successful peer.
In the latter case,
  individuals adjust their strategies
  based on a comparison of their payoffs from the evolutionary game to a value they aspire,
  called the level of aspiration.
Unlike imitation processes of pairwise comparison,
  aspiration-driven updates do not require additional information about the strategic environment
  and can thus be interpreted as being more spontaneous.
Recent work has mainly focused on understanding
  how aspiration dynamics alter the evolutionary outcome in structured populations.
However,
  the baseline case for understanding strategy selection
  is the well-mixed population case,
  which is still lacking sufficient understanding.
We explore
  how aspiration-driven strategy-update dynamics under imperfect rationality
  influence the average abundance of a strategy in multi-player evolutionary games with two strategies.
We analytically derive a condition
  under which a strategy is more abundant than the other in the weak selection limiting case.
This approach has a long standing history in evolutionary game
  and is mostly applied for its mathematical approachability.
Hence,
  we also explore strong selection numerically,
  which shows that our weak selection condition
  is a robust predictor of the average abundance of a strategy.
The condition turns out to differ from that of a wide class of imitation dynamics,
  as long as the game is not dyadic.
Therefore a strategy favored under imitation dynamics
  can be disfavored under aspiration dynamics.
This does not require any population structure
  thus highlights the intrinsic difference between imitation and aspiration dynamics.
\end{abstract}
\newpage

\onehalfspacing

\section{Introduction}
\label{sec:Intro}

In the study of population dynamics,
  it turns out to be very useful to classify individual interactions
  in terms of evolutionary games \cite{sigmund:book:2010}.
Early mathematical theories of strategic interactions
  were based on the assumption of rational choice
  \cite{neumann:book:1944,nash:PNAS:1950}:
  an agent's optimal action depends on its expectations on the actions of others,
  and each of the other agents' actions depend on their expectations about the focal agent.
In evolutionary game theory,
  successful strategies spread by reproduction or imitation in a population
  \cite{Smith1973Nature,Weibull1995MIT,Hofbauer1998Cambridge,gintis:book:2000,Nowak2006Harvard}.

Evolutionary game theory not only provides a platform for explaining
  biological problems of frequency dependent fitness
  and complex individual interactions such as cooperation and coordination
  \cite{Nowak2004Science,Imhof2006JMB}.
In finite populations,
  it also links the neutral process of evolution \cite{kimura:book:1983}
  to frequency dependence by introducing an intensity of selection
  \cite{Taylor2004BMB,altrock:PRE:2009,hilbe:BMB:2011,Arnoldt2012Interface}.
Evolutionary game theory can also be used to study cultural dynamics
  including human strategic behavior and updating
  \cite{bendor:PNAS:1995,Milinski2006PNAS,Traulsen2010PNAS}.
One of the most interesting open questions is
  How do individuals update their strategies
  based on the knowledge and conception of others and themselves?

Two fundamentally different mechanisms can be used
  to classify strategy updating and population dynamics
  based on individuals' knowledge about their strategic environment or themselves:
  imitation of others and self-learning based on one's own aspiration.
In imitation dynamics,
  players update their strategies after a comparison
  between their own and another individual's success in the evolutionary game
  \cite{Traulsen2007JTB,traulsen:PRL:2005,Wu2010PRE}.
For aspiration-driven updating,
  players switch strategies if an aspiration level is not met,
  where the level of aspiration is an intrinsic property of the focal individual
  \cite{Szabo1998PRE,Roca2011PNAS,Szabo2007PR,Chen2008PRE}.
In both dynamics,
  novel strategies cannot emerge without additional mechanisms
  such as spontaneous exploration of strategy space (similar to mutation)
  \cite{Fudenberg2006JET,Lessard2007JMB,Traulsen2007JTB,traulsen:PNAS:2009,Tarnita2009JTBa,Rand2012Nature}.
The major difference is that
  the latter does not require any knowledge about the payoffs of others.
Thus aspiration level based dynamics,
  a form of self-learning,
  require less information about an individual's strategic environment than imitation dynamics.

Aspiration-driven strategy-update dynamics
  are commonly observed in studies of animal and human behavioral ecology.
For example, fish would ignore social information
  when they have relevant personal information \cite{Bergen2004PRSB},
  and experienced ants hunt for food
  based on their own previous chemical trials rather than imitating others \cite{Grueter2011BES}.
Furthermore,
  a form of aspiration-level-driven dynamics
  play a key role in the individual behaviors in rat populations \cite{Galef2008AB}.
These examples clearly show that
  the idea behind aspiration dynamics, i.e., self-evaluation,
  is present in the animal world.
In behavioral sciences,
  such aspiration-driven strategy adjustments generally operate on the behavioral level.
However, it can be speculated that
  self-learning processes can have such an effect that
  it might actually have a downward impact on regulatory,
  and thus genetic levels of brain and nervous system.
This, in turn,
  could be seen as a mechanism that
  alters the rate of genetic change \cite{Hoppitt2013Princeton}.
Whereas such wide reaching systemic alterations are more speculative,
  it is clear that
  aspiration levels play a role in human strategy updating \cite{Roca2011PNAS}.

We study the statistical mechanics of a simple case of
  aspiration-driven self-learning dynamics in well-mixed populations of finite size.
Deterministic and stochastic models of imitation dynamics
  have been well studied in both well-mixed and structured populations
  \cite{Hofbauer1998Cambridge,Nowak2004Nature,Fudenberg2006JET,Szabo2007PR,Traulsen2007JTB,Gokhale2010PNAS}.
For aspiration dynamics,
  numerous works have emerged studying population dynamics on graphs,
  but its impact in well-mixed populations--a basic reference case, one would think--is far less well understood.
Although deterministic aspiration dynamics,
  i.e., a kind of win-stay-lose-shift dynamics,
  in which individuals are perfectly rational
  have been analyzed \cite{Posch1999PRSB},
  it is not clear how processes with imperfect rationality influence the evolutionary outcome.
Here, we ask whether a strategy favored under pairwise comparison driven imitation dynamics
  can become disfavored under aspiration-driven self-learning dynamics.
To this end,
  in our analytical analysis,
  we limit ourselves to the weak selection,
  or weak rationality approximation,
  where payoffs via the game play little role in the decision-making \cite{Nowak2004Nature}.
As it has been shown that under weak selection,
  the favored strategy is invariant for a wide class of imitation processes
  \cite{Lessard2007JMB,Wu2010PRE,Lessard2011DGA}.
We show that for pairwise games,
  the aspiration dynamics and the imitation dynamics always share the same favored strategies.
For multi-player games,
  however, the weak selection criterion under aspiration dynamics
  that determines whether a strategy is more abundant than the other
  differs from the criterion under imitation dynamics.
This paves the way to construct multi-player games,
  for which aspiration dynamics favor one strategy
  whereas imitation dynamics favor another.
Furthermore, in contrast to deterministic aspiration dynamics,
  if the favored strategy is determined by a global aspiration level,
  the average abundance of a strategy in the stochastic aspiration dynamics
  is invariant with respect to the aspiration level,
  provided selection is weak.
We also extrapolate our results to stronger selection cases through numerical simulation.

\section{Mathematical Model}

\label{sec:Model}
\setcounter{equation}{0}

\subsection{Evolutionary games}

We consider evolutionary game dynamics with two strategies and $d$ players.
From these,
  the more widely studied $2\times 2$ games
  emerge as a special case \cite{Gokhale2010PNAS}.
In individual encounters,
  players obtain their payoffs from simultaneous actions.
A focal player can be of type $A$, or $B$,
  and encounter a group containing $k$ other players of type $A$,
  to receive the payoff $a_k$, or $b_k$.
For example,
  a $B$ player,
  which encounters $d-1$ individuals of type $A$,
  obtains payoff $b_{d-1}$.
An $A$ player in a group of one other $A$ player and thus $d-2$ $B$ players
  obtains payoff $a_1$.
All possible payoffs of a focal individual
  are uniquely defined by the number of $A$ in the group,
  such that the payoff matrix reads
\begin{align}
\label{Payoff_Matrix}
\begin{tabular}{ l | c c c c c c}
    & $d-1$     & $\dots$ & $k$     & $\dots$ & $0$     \\
\hline
$A$ & $a_{d-1}$ & $\dots$ & $a_{k}$ & $\dots$ & $a_{0}$ \\
$B$ & $b_{d-1}$ & $\dots$ & $b_{k}$ & $\dots$ & $b_{0}$ \\
\end{tabular}
\end{align}
For any group engaging in a one-shot game,
  we can obtain each member's payoff according to this matrix.

In a finite well-mixed population of size $N$,
  groups of size $d$ are assembled randomly,
  such that the probability of choosing a group
  that consists of another $k$ players of type $A$,
  and of $d-1-k$ players of type $B$,
  is given by a hypergeometric distribution \cite{Graham1994AddisonWesley}.
For example,
  the probability that an $A$ player is in a group of $k$ other $A$s
  is given by
  $\text{prob}_A (\,k \,|\, N,i,d\,)=(\,C_{i-1}^{k}\,C_{N-i}^{d-1-k}\,)/C_{N-1}^{d-1}$,
  where $i$ ($i\geq d$) is the number of $A$ players in the population,
  and $C^k_n=n!/(\,k!\,(n-k)!\,)$ is the binomial coefficient.

\begin{figure}[!ht]
\begin{center}
\includegraphics[width=0.95\textwidth]{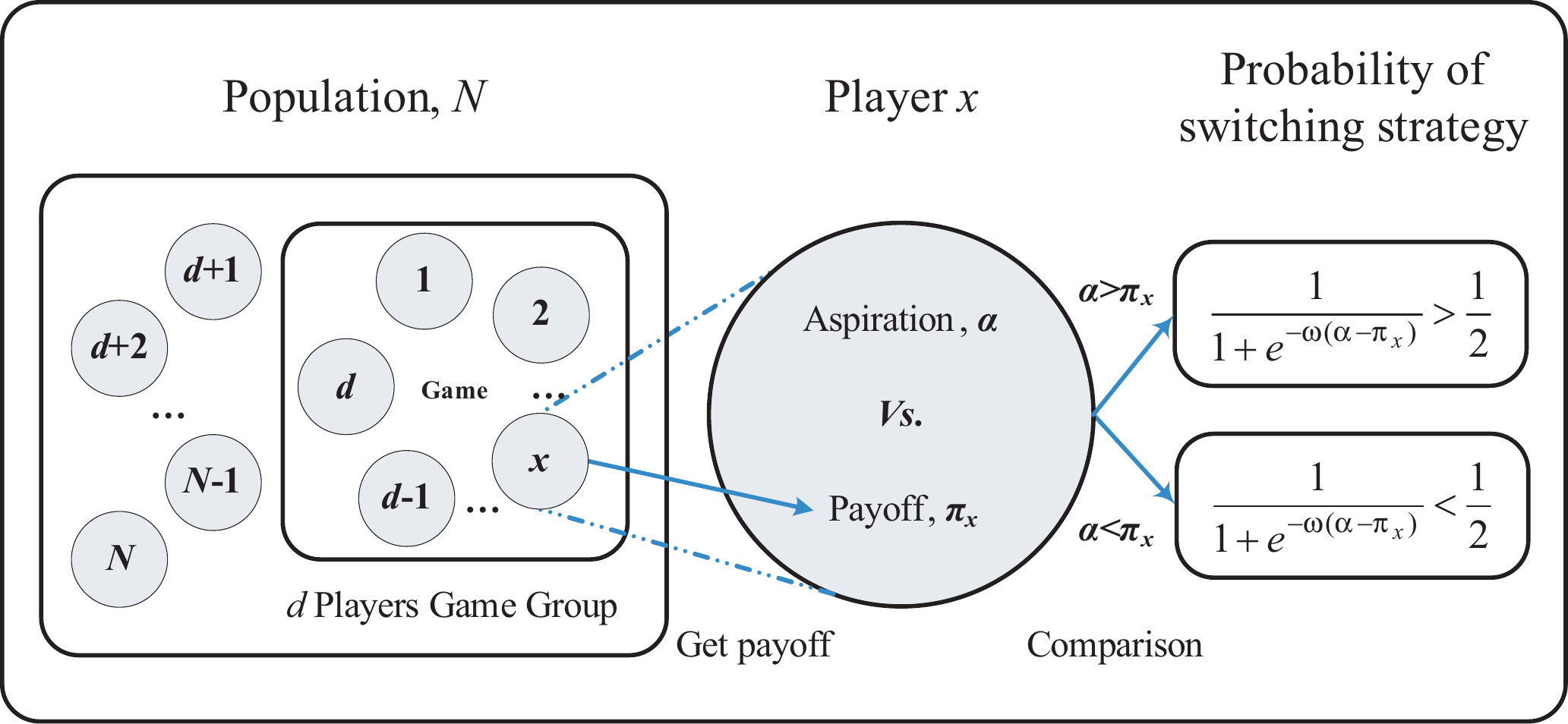}
\end{center}
\caption{
\small
{\bf Evolutionary game dynamics of $d$-player interactions driven by global aspiration.}
In our mathematical model of human strategy updating driven by self-learning,
  a group of $d$ players is chosen randomly from the finite population to play the game.
According to this,
  game players calculate and obtain their actual payoffs.
They are more likely to stochastically switch strategies
  if the payoffs they aspire are not met.
On the other hand,
  the higher the payoffs compared to the aspiration level $\alpha$ are,
  the less likely they switch their strategies.
Besides, strategy switching is also determined by a selection intensity $\omega$.
For vanishing selection intensity,
  switching is entirely random irrespective of payoffs and the aspiration level.
For increasing selection intensity,
  the self-learning process becomes increasingly more ``optimal''
  in the sense that for high $\omega$,
  individuals tend to always switch when they are dissatisfied,
  and never switch when they are tolerant.
We examine the simplest possible setup,
  where the level of aspired payoff $\alpha$ is a global parameter
  that does not change with the dynamics.
We show that, however,
  statements about the average abundance of a strategy
  do not depend on $\alpha$ under weak selection ($\omega\ll1$).
}
\label{Model}
\end{figure}

The expected payoffs for any $A$ or $B$ in a population of size $N$,
  with $i$ players of type $A$ and $N-i$ players of type $B$,
  are given by
\begin{align}
\pi_{A}(i)&=\sum_{k=0}^{d-1}\frac{C_{i-1}^{k}\,C_{N-i}^{d-1-k}}{C_{N-1}^{d-1}}\,a_{k},
\label{Payoff_C} \\
\pi_{B}(i)&=\sum_{k=0}^{d-1}\frac{C_{i}^{k}\,C_{N-i-1}^{d-1-k}}{C_{N-1}^{d-1}}\,b_{k}.
\label{Payoff_D}
\end{align}
In summary,
  we define a $d$-player stage game \cite{gintis:book:2000},
  shown in Eq.~(\ref{Payoff_Matrix}),
  from which the evolutionary game emerges
  such that each individual obtains an expected payoff
  based on the current composition of the well-mixed population.
In the following,
  we introduce an update rule based on a global level of aspiration.
This allows us to define a Markov chain
  describing the inherently stochastic dynamics in a finite population:
  probabilistic change of the composition of the population
  is driven by the fact that
  each individual compares its actual payoff to an imaginary value that it aspires.
Note here that
  we are only interested in the simplest way to model such a complex problem
  and do not address any learning process
  that may adjust such an aspiration level as the system evolves.
For a sketch of the aspiration-driven evolutionary game,
  see Fig.~\ref{Model}.

\subsection{Aspiration-level driven stochastic dynamics}

In addition to the inherent stochasticity in finite populations,
  there is randomness in the process of individual assessments of one's own payoff
  as compared to a random sample of the rest of the population;
  even if an individual knew exactly what to do,
  he might still fail to switch to an optimal strategy,
  e.g., due to a trembling hand
  \cite{selten:IJGT:1975,myerson:IJGT:1978}.

Here we examine the simplest case
  of an entire population having a certain level of aspiration.
Players needn't see any particular payoffs but their own,
  which they compare to an aspired value.
This level of aspiration, $\alpha$,
  is a variable that influences the stochastic strategy updating.
The probability of switching strategy is random
  when individuals' payoffs are close to their level of aspiration,
  reflecting the basic degree of uncertainty in the population.
When payoffs exceed the aspiration,
  strategy switching is unlikely.
At high values of aspiration compared to payoffs,
  switching probabilities are high.

The level of aspiration
  provides a global benchmark of tolerance or dissatisfaction in the population.
In addition,
  when modeling human strategy updating,
  one typically introduces another global parameter
  that provides a measure for
  how important individuals deem the impact of the actual game played on their update,
  the intensity of selection, $\omega$.
Irrespective of the aspiration level and the frequency dependent payoff distribution,
  vanishing values of $\omega$ refer to nearly random strategy updating.
For large values of $\omega$,
  individuals' deviations from their aspiration level
  have a strong impact on the dynamics.

Note that
  although the level of aspiration
  is a global variable and does not differ individually,
  due to payoff inhomogeneity
  there can always be a part of the population
  that seeks to switch more often
  due to dissatisfaction with the payoff distribution.

In our microscopic update process,
  we randomly choose an individual, $x$, from the population,
  and assume that the payoff of the focal individual is $\pi_x$.
To model stochastic self-learning of aspiration-driven switching,
  we can use the following probability function
\begin{align}
g(\,\alpha-\pi_x\,) = \frac{1}{1+e^{-\omega\,(\,\alpha-\pi_{x}\,)}},
\label{Updating_Rule}
\end{align}
  which is similar to the Fermi-rule
  \cite{blume:GEB:1993,Szabo1998PRE},
  but replaces a randomly drawn opponent's payoff by one's own aspiration.
The wider the positive gap between aspiration and payoff,
  the higher the switching probability.
Reversely,
  if payoffs exceed the level of aspiration
  individuals become less active with increasing payoffs.
The aspiration level, $\alpha$,
  provides the benchmark used to evaluate
  how ``greedy'' an individual is.
Higher aspiration levels mean that
  individuals aspire to higher payoffs.
If payoffs meet aspiration,
  individuals remain random in their updates.
If payoffs are below aspiration,
  switching occurs with probability larger than random;
  if they are above aspiration,
  switching occurs with probability lower than random.
The selection intensity governs
  how strict individuals are in this respect.
For $\omega=0$,
  strategy switching is entirely random (neutral).
Low values of $\omega$
  lead to switching only slightly different from random
  but follow the impact of $\alpha$.
For increasing $\omega$,
  the impact of the difference between payoffs and the aspiration
  becomes more important.
In the case of $\omega\to\infty$,
  individuals are strict in the sense that
  they either switch strategies with probability one
  if they are not satisfied,
  or stay with their current strategy
  if their aspiration level is met or overshot.

The spread of successful strategies
  is modeled by a birth and death process in discrete time.
In one time step,
  three events are possible:
  the abundance of $A$, $i$,
  can increase by one with probability $p(\,i\to i+1\,)=T_i^+$,
  decrease by one with probability $p(\,i\to i-1\,)=T_i^-$,
  or stay the same with probability $p(\,i\to i\,)=T_i^0$.
All other transitions occur with probability zero.
The transition probabilities are given by
\begin{align}
T_{i}^{+}&=\frac{N-i}{N}\,\frac{1}{1+e^{-\omega\,[\,\alpha-\pi_{B}(i)\,]}},
\label{TC_i+1} \\
T_{i}^{-}&=\frac{i}{N}\, \frac{1}{1+e^{-\omega\,[\,\alpha-\pi_{A}(i)\,]}},
\label{TC_i-1}\\
T_{i}^{0}&=1-T_{i}^{+}-T_{i}^{-}
\label{TC_i}.
\end{align}
In each time step,
  a randomly chosen individual evaluates its success in the evolutionary game,
  given by Eqs.~(\ref{Payoff_C}), (\ref{Payoff_D}),
  compares it to the level of aspiration,
  and then changes strategy with probability lower than $1/2$
  if its payoff exceeds the aspiration.
Otherwise, it switches with probability greater than $1/2$,
  except when the aspiration level is exactly met,
  in which case it switches randomly
  (note that this is very unlikely to ever be the case).

Compared to imitation (pairwise comparison) dynamics,
  our self-learning process,
  which is essentially an Ehrenfest-like Markov chain,
  has some different characteristics.
Without the introduction of mutation or random strategy exploration,
  there exists a stationary distribution for the aspiration-driven dynamics.
Even in a homogenous population,
  there is a positive probability that
  an individual can switch to another strategy
  due to the dissatisfaction resulting from payoff-aspiration difference.
This facilitates the escape from the states
  that are absorbing in the pairwise comparison process
  and other Moran-like evolutionary dynamics.
Hence there exists a nontrivial stationary distribution of the Markov chain
  satisfying detailed balance.
Specifically, for the case of $\omega=0$ (neutral selection),
  the dynamics defined by Eqs.~(\ref{TC_i+1})--(\ref{TC_i})
  are characterized by linear rates,
  while these rates are quadratic
  for the neutral imitation dynamics and Moran process.

In the following analysis and discussion,
  we are interested in the limit of weak selection, $\omega\ll1$,
  and its ability to aptly predict the success of cooperation
  in commonly used evolutionary $d-$player games.
The limit of weak selection,
  which has a long standing history
  in population genetics and molecular evolution \cite{kimura:book:1983}
  also plays a role in social learning and cultural evolution.
Recent experimental results suggest that
  the intensity with which human subjects adjust their strategies
  might be low \cite{Traulsen2010PNAS}.
Although it has been unclear to what degree and in what way
  human strategy updating deviates from random
  \cite{helbing:PNAS:2010,grujic:PLoSOne:2012},
  the weak selection limit
  is of importance to quantitatively characterize the evolutionary dynamics.
In the limiting case of weak selection,
  we are able to analytically classify strategies
  with respect to the neutral benchmark, $\omega\to0$
  \cite{Nowak2004Nature,Traulsen2007JTB,altrock:NJP:2009,Wu2010PRE,taylor:JTB:2006}.
We note that
  a strategy is favored by selection,
  if its average equilibrium frequency under weak selection
  is greater than one half.
In order to come to such a quantitative observation,
  we need to calculate the stationary distribution over frequencies of strategy $A$.

\subsection{Stationary distribution}

The Markov chain given by Eqs.~(\ref{TC_i+1})--(\ref{TC_i})
  is a one dimensional birth and death process with reflecting boundaries.
It satisfies the detailed balance condition
  $\psi_{j-1}\,T_{j-1}^{+}=\psi_{j}\,T_{j}^{-}$,
  where \[(\,\psi_0,\psi_1,\dots,\psi_{j},\dots,\psi_N\,)\]
  is the stationary distribution over frequencies of $A$ in equilibrium
  \cite{Kampen2007Elsevier,Gardiner2004Springer}.
Considering $\sum_{j=0}^{N}\psi_{j}=1$,
  we find the exact solution by recursion,
  given by
\begin{align}
\psi_{j}=
\begin{cases}
j=0:\,\,\,\frac{1}{ 1+\sum_{k=1}^{N} q_{0\,k-1}^{+}/q_{1\,k}^{-} }\\
\\
j>0:\,\,\,\frac{ q_{0\,j-1}^{+}/q_{1\,j}^{-} }{ 1+\sum_{k=1}^{N} q_{0\,k-1}^{+}/q_{1\,k}^{-} }
\end{cases},
\label{Stationary Distribution}
\end{align}
  where $q_{j\,k}^\pm=\prod_{l=j}^{k} T_l^{\pm}$
  is the probability of successive transitions from $j$ to $k$.
The analytical solution Eq.~(\ref{Stationary Distribution})
  allows us to find the exact value of the average abundance of strategy $A$,
\begin{align}
\langle X_A \rangle(\omega)=\sum_{j=0}^{N} \frac{j}{N}\,\psi_{j}(\omega),
\label{Average Abundance}
\end{align}
  for any strength of selection.

\section{Results and Discussion}
\label{sec:Res}
\setcounter{equation}{0}

It has been shown that
  imitation processes are similar to each other under weak selection
  \cite{Lessard2007JMB,Wu2010PRE,Lessard2011DGA}.
Thus in order to compare the essential differences
  between imitation processes and aspiration process,
  we consider such selection limit.
To better understand the effects of
  selection intensity, aspiration level, and payoff matrix
  on the average abundance of strategy $A$,
  we further analyze which strategy is more abundant
  based on Eq.~(\ref{Stationary Distribution}).
For a fixed population size,
  under weak selection, i.e. $\omega\rightarrow0$,
  the stationary distribution $\psi_{j}(\omega)$
  can be expressed approximately as
\begin{align}
\psi_{j}(\omega)\approx \psi_{j}(0)+\omega\,\left[\,\frac{\partial}{\partial\,\omega}\,\psi_{j}(\omega)\,\right]_{\omega=0},
\label{Stationary Distribution Approximation}
\end{align}
  where the neutral stationary distribution is simply given by
  $\psi_{j}(0)=C_{N}^{j}/2^{N}$,
  and the first order term of this Taylor expansion amounts to
\begin{align}
\left[\,\frac{\partial}{\partial\,\omega}\,\psi_{j}(\omega)\,\right]_{\omega=0}
=\frac{C_{N}^{j}}{2^{N+1}}\,
\left\{
\sum\limits_{k=1}^{j}[\,\pi_{A}(k)-\pi_{B}(k-1)\,]
-\frac{1}{2^{N}}\,\sum\limits_{k=1}^{N}\,C_{N}^{k}\sum\limits_{l=1}^{k}[\,\pi_{A}(l)-\pi_{B}(l-1)\,]
\right\}.
\label{psi'_j(0)}
\end{align}
Interestingly, in the limiting case of weak selection,
  the first order approximation of the stationary distribution of $A$
  does not depend on the aspiration level.
For higher order terms of selection intensity,
  however, $\psi_{j}(\omega)$ does depend on the aspiration level.

In the following
  we discuss the condition under which a strategy is favored
  and compare the predictions for stationary strategy abundance
  under self-learning and under imitation dynamics.
Thereafter we consider three prominent examples of games with multiple players
  through analytical, numerical and simulation methods,
  the results of which are detailed
  in Figs.~\ref{fig:PGG}--\ref{fig:SDG} and Appendix~\ref{Appendix II}.
All three examples are social dilemmas in the sense that
  the Nash equilibrium of the one-shot game is not the social optimum.
First, the widely studied public goods game
  represents the class of games
  where there is only one pure Nash equilibrium \cite{axelrod:book:1984}.
Next, the public goods game with a threshold,
  a simplified version of the collective risk dilemma
  \cite{Milinski2008PNAS,santos:PNAS:2011,hilbe:PlosOne:2013},
  represents the class of coordination games with multiple pure Nash equilibria,
  depending on the threshold.
Last, we consider the $d$-player volunteer's dilemma, or snowdrift game,
  which has a mixed Nash equilibrium
  \cite{hauert:Nature:2004,doebeli:Science:2004}.

\subsection{Average abundance of strategy $A$}

Based on the approximation (\ref{Stationary Distribution Approximation}),
  for any symmetric multi-player game
  with two strategies of normal form (\ref{Payoff_Matrix}),
  we can now calculate a weak selection condition
  such that in equilibrium $A$ is more abundant than $B$.
Since for neutrality,
  $\psi_{j}(0)=C_{N}^{j} / 2^{N}$ holds
  and thus $\langle X_A \rangle(0)=1/2$,
  it is sufficient to consider positivity of the sum of
  $j\omega\,[\,\partial_\omega\,\psi_{j}(\omega)\,]_{\omega=0}/N$
  over all $j=0,\dots,N$.
Under weak selection,
  strategy $A$ is favored by selection,
  i.e., $\langle X_A \rangle(\omega)>1/2$,
  if
\begin{align}
\sum_{k=0}^{d-1} C_{d-1}^{k}\left(\,a_{k}-b_{k}\,\right)>0,
\label{Criterion Condition}
\end{align}
  which holds for any $d$-player games with two strategies
  in a population with more than two individuals.
For a detailed derivation of our main analytical result,
  see Appendix \ref{Appendix}.
Note that for a two-player game, $d=2$,
  the above condition simplifies to
  $a_{1}+a_{0}>b_{1}+b_{0}$,
  which is similar to the concept of risk-dominance
  translated to finite populations \cite{Nowak2004Nature}.

The left hand side expression of inequality (\ref{Criterion Condition})
  can also be compared to a similar condition
  under the class of pairwise comparison processes
  \cite{Szabo1998PRE,Traulsen2007JTB},
  where two randomly selected individuals compare their payoffs
  and switch with a certain probability based on the observed inequality.
Typically, weak selection results for pairwise comparison processes
  lead to the result that
  strategy $A$ is favored by selection if
  \cite{Huberman1993PNAS,Nowak1994PNAS,Nowak2004Nature}
\begin{align}
\sum_{k=0}^{d-1} \left(\,a_{k}-b_{k}\,\right)>0,
\label{PairwiseComp_Condition}
\end{align}
  which applies both,
  to evaluate whether fixation of $A$ is more likely than fixation of $B$,
  or whether the average abundance of $A$ is greater than one half
  under weak mutation and weak selection,
  that can be shown using properties of the embedded Markov chain \cite{allen:JMB:2012}.
The sums on the left hand sides of
  (\ref{Criterion Condition}) and (\ref{PairwiseComp_Condition})
  can thus be compared with each other
  in order to reveal the nature
  of our self-learning process driven by a global aspiration level.

Our main result, Eq.~(\ref{Criterion Condition}),
  holds for a variety of self-learning dynamics,
  not only for the probability function given by Eq.~(\ref{Updating_Rule}).
Considering the general self-learning function
  $g[\,\omega\,(\,\alpha-\pi_x\,)\,]$ with $g(0)\neq0$,
  here $g(x)$ is strictly increasing with increasing $x$.
Denoting $u=\omega\,(\,\alpha-\pi_{x}\,)$,
  we have $g'(\omega)=g'(u)\, u'(\omega)$.
Then, for $\omega\rightarrow0$,
  $\left.g'(\omega)\right|_{\omega=0}=g'(0)\,(\,\alpha-\pi_{x}\,)$,
  and Eq.~(\ref{psi'_j(0)}) can be rewritten in a more general form
\begin{align}
\left[\,\frac{\partial}{\partial\,\omega}\,\psi_{j}(\omega)\,\right]_{\omega=0}
=\frac{g'(0)}{g(0)}\,\frac{C_{N}^{j}}{2^{2N}}\,
\left\{
2^{N}\,\sum_{k=1}^{j}[\,\pi_{A}(k)-\pi_{B}(k-1)\,]
-\sum_{k=1}^{N}\,C_{N}^{k}\sum_{i=1}^{k}[\,\pi_{A}(i)-\pi_{B}(i-1)\,]
\right\}.
\label{General_psi'_j(0)}
\end{align}
Since $g'(0)/g(0)$ is a positive constant,
  Eq.~(\ref{Criterion Condition}) is still valid
  for any such probability function $g(x)$,
  see Appendix~\ref{Appendix}.

\subsection{Linear public goods game}

\begin{figure}[!ht]
\begin{center}
\includegraphics[width=0.95\textwidth]{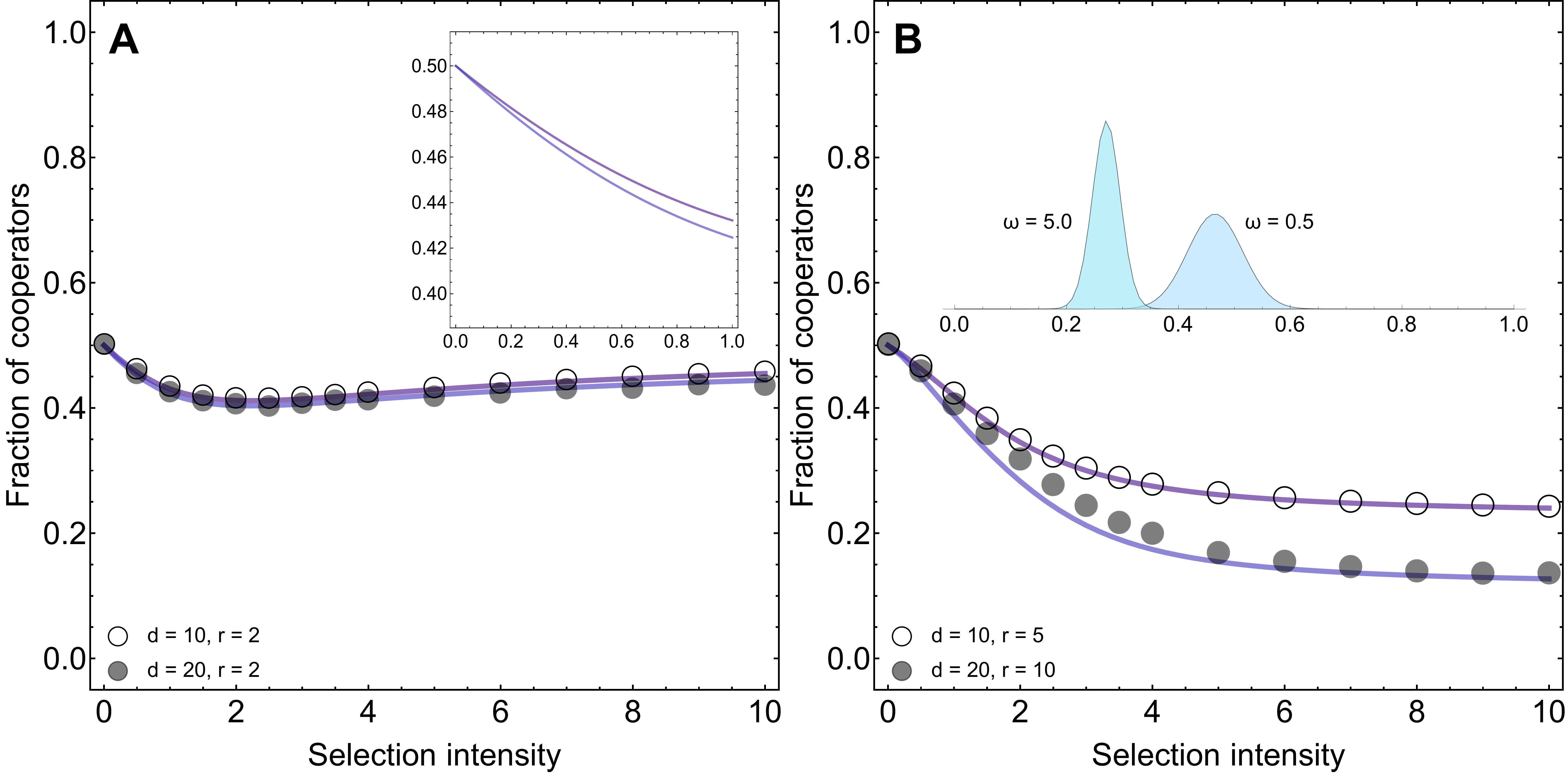}
\caption{
\small
{\bf Mean (stationary) fraction of cooperators for the linear public goods game.}
The common parameters are
  aspiration level $\alpha=1$, population size $N=100$, and cost of cooperation $c=1$.
In both panels,
  the group sizes are $d=10$ (dark shaded), and $d=20$ (light shaded).
Panel {\bf A} shows the mean fraction of cooperators
  as a function of selection intensity for $r=2$,
  the inset shows a detail for lower selection intensities.
Panel {\bf B} shows the mean fraction of cooperators
  as a function of selection intensity for $r=d/2$.
The inset shows the stationary distribution for $d=10$, $r=5$ and $\omega=0.5,\,5$.
}\label{fig:PGG}
\end{center}
\end{figure}

Public goods games emerge
  when groups of players engage in the sustenance of common goods.
Cooperators $A$ pay an individual cost in form of a contribution $c$
  that is pooled into the common pot.
Defectors $B$ do not contribute.
The pot is then multiplied by a characteristic multiplication factor $r$
  and shared equally among all individuals in the group,
  irrespective of contribution.
If the multiplication factor is smaller than the size of the group $d$,
  each cooperator recovers only a fraction of the initial investment.
Switching to defection would always be beneficial
  in a pairwise comparison of the two strategies.
The payoff matrix thus reads
\begin{align}
\label{PPG_Matrix}
\begin{tabular}{ l | c c c c c c c }
    & $d-1$                 & $\dots$ & $k$                     & $\dots$ & $1$                   & $0$                \\
\hline
$A$ & $r\,c-c$              & $\dots$ & $\frac{k+1}{d}\,r\,c-c$ & $\dots$ & $\frac{2}{d}\,r\,c-c$ & $\frac{r\,c}{d}-c$ \\
$B$ & $\frac{d-1}{d}\,r\,c$ & $\dots$ & $\frac{k}{d}\,r\,c$     & $\dots$ & $\frac{r\,c}{d}$      & $0$                \\
\end{tabular}
\end{align}
  where $1<r<d$ is typically assumed.
Since $a_{k}-b_{k}=c\,(r/d-1)$ is a negative constant
  for any number of cooperators in the group,
  we find that
\begin{align}
\sum_{k=0}^{d-1} C_{d-1}^{k}\,(\,a_{k}-b_{k}\,)=2^{d-1}\,c\,\left(\,\frac{r}{d}-1\,\right)
\end{align}
  is always negative.
Cooperation cannot be the more abundant strategy
  in the well-mixed population (see Fig.~\ref{fig:PGG}).
However, if the self-learning dynamics
  are driven by a sufficiently high aspiration level,
  individuals are constantly dissatisfied and
  switch strategy frequently, even as defectors,
  such that cooperation can break even if selection is strong enough,
  namely $\lim_{\alpha\to\infty}\langle X_A \rangle=1/2$
  for all values $\omega$.
On the other hand,
  if the aspiration level is low,
  cooperators switch more often than defectors
  such that the average abundance of $A$
  assumes a value closer to the evolutionary stable state of full defection,
  which depends on $\omega$.
In the extreme case of very low $\alpha$ and strong selection,
  defectors fully dominate,
  thus the stationary measure retracts to the all defection state.

\subsection{Threshold public goods game}
\begin{figure}[!ht]
\begin{center}
\includegraphics[width=0.95\textwidth]{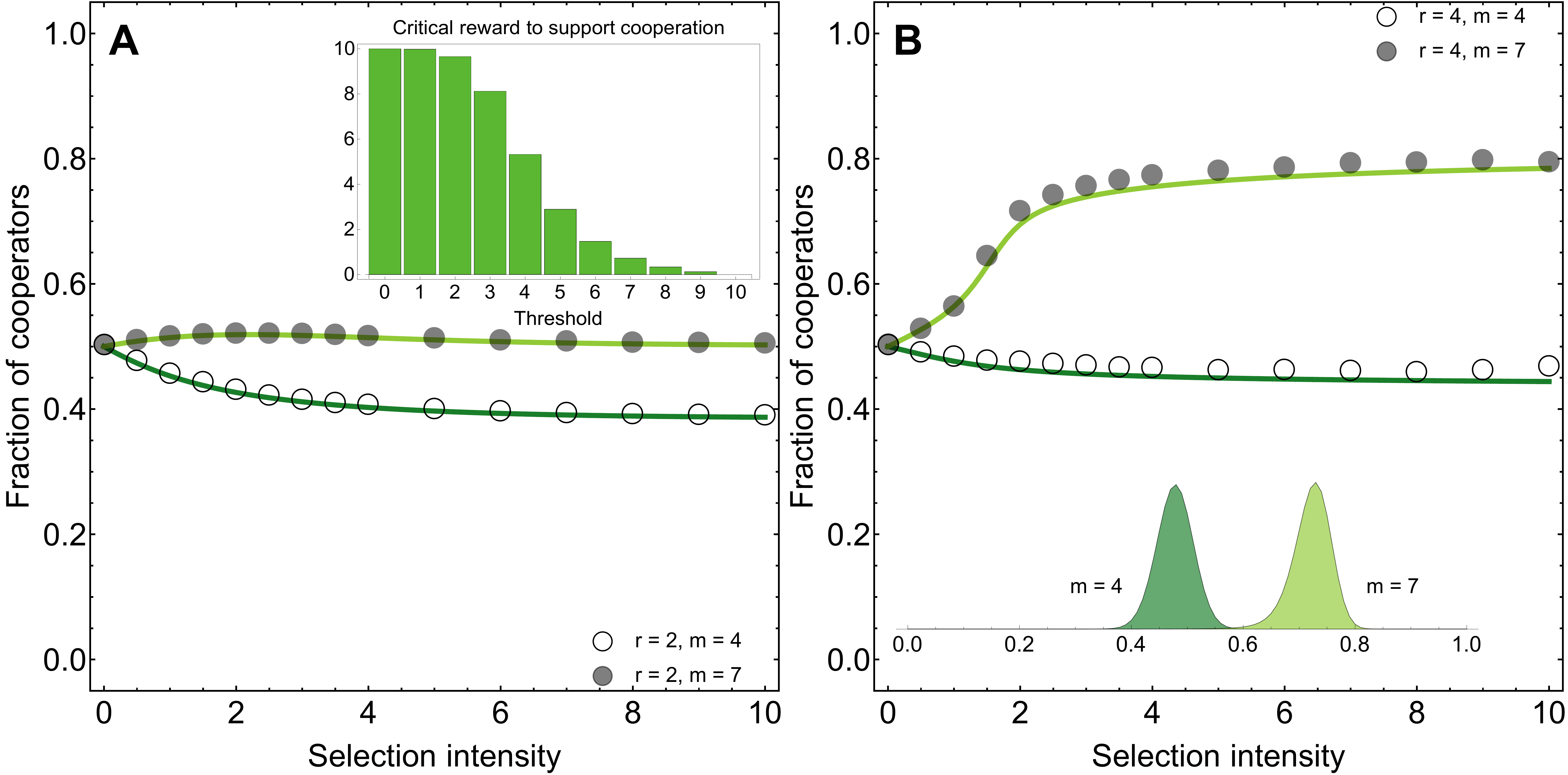}
\caption{
\small
{\bf Mean (stationary) fraction of cooperators for the threshold public goods game.}
The common parameters are
  aspiration level $\alpha=1$, population size $N=100$, group size $d=10$, and cost of cooperation $c=1$.
In both panels,
  the threshold sizes are $m=4$ (dark shaded), and $m=7$ (light shaded).
Panel {\bf A} shows the mean fraction of cooperators
  as a function of selection intensity for $r=2$.
The inset shows the critical multiplication factor
  above which cooperation is more abundant as a function of the threshold $m$.
High thresholds lower the critical multiplication factor of the public good
  such that cooperation can become more abundant than defection.
Panel {\bf B} shows the mean fraction of cooperators
  as a function of selection intensity for $r=4$.
The inset shows the stationary distribution for $d=10$, $r=4$, $\omega=2$, and $m=4,\,7$.
}\label{fig:TPGG}
\end{center}
\end{figure}

Here we consider the following public goods game with a threshold
  in the sense that the good becomes strictly unavailable
  when the number of cooperators in a group
  is below a critical threshold, $m$.
This threshold becomes a new strategic variable.
Here, $c$ is an initial endowment given to each player,
  which is invested in full by cooperators.
Whatever the cooperators manage to invest
  is multiplied by $r$
  and redistributed among all players in the group irrespective of strategy,
  if the threshold investment $m\,c$ is met.
Defectors do not make any investment,
  and thus have an additional payoff of $c$,
  as long as the threshold is met.
Once the number of cooperators is below $m$,
  all payoffs are zero,
  which compares to the highest risk possible
  (loss of endowment and investment with certainty)
  in what is called the collective-risk dilemma
  \cite{Milinski2008PNAS,hilbe:PlosOne:2013}.
The payoff matrix for the two strategies,
  cooperation $A$, and defection $B$,
  reads
\begin{align}
\label{TPPG_Matrix}
\begin{tabular}{ l | c c c c c c c c}
    & $d-1$                   & $\dots$ & $m$                   & $m-1$               & $m-2$ & $\dots$ & $0$ \\
\hline
$A$ & $r\,c$                  & $\dots$ & $\frac{m+1}{d}\,r\,c$ & $\frac{m}{d}\,r\,c$ & $0$   & $\dots$ & $0$ \\
$B$ & $\frac{d-1}{d}\,r\,c+c$ & $\dots$ & $\frac{m}{d}\,r\,c+c$ & $0$                 & $0$   & $\dots$ & $0$ \\
\end{tabular}
\end{align}
We can examine
  when the self-learning process favors cooperation.
We can also seek to make a statement about
  whether under self-learning dynamics,
  cooperation performs better than under pairwise comparison process.
For self-learning dynamics,
  we find
  \[\sum_{k=0}^{d-1}C_{d-1}^{k}\,(\,a_{k}-b_{k}\,)=\sum_{k=m}^{d-1}C_{d-1}^{k}\,(\,r\,c/d-c\,)+C_{d-1}^{m-1}\,(\,m\,r\,c/d\,)\]
  while the equivalent statement for pairwise comparison processes
  based on the same payoff matrix would be
  $\sum_{k=0}^{d-1}(\,a_{k}-b_{k}\,)=c\,(\,r+m-d\,)$.
Thus, the criterion of self-learning dynamics
  can be written as
  \[r>(\,d\,\sum_{k=m}^{d-1}C_{d-1}^{k})/(\,\sum_{k=m}^{d-1}C_{d-1}^{k}+m\,C_{d-1}^{m-1}\,)\]
  whereas positivity of the imitation processes condition, $\sum_{k=0}^{d-1}(\,a_{k}-b_{k}\,)>0$, simply leads to $r>d-m$.
Comparing the two conditions,
  we find
\begin{align}
\label{TPGG_03}
(\,d-m\,)-\frac{d\,\sum_{k=m}^{d-1}C_{d-1}^{k}}{\sum_{k=m}^{d-1}C_{d-1}^{k}+m\,C_{d-1}^{m-1}}
=\frac{m}{\sum_{k=m}^{d-1}C_{d-1}^{k}+m\,C_{d-1}^{m-1}}\,\Gamma_{d,m}.
\end{align}
Since the first factor on the right hand side of Eq.~(\ref{TPGG_03})
  is always positive,
  the factor
\begin{align}
\label{TPGG_04}
\Gamma_{d,m} = (\,d-m\,)\,C_{d-1}^{m-1}-\sum_{k=m}^{d-1}C_{d-1}^{k}
\end{align}
  determines the relationship
  between self-learning dynamics and pairwise comparison processes:
  for sufficiently large threshold $m$,
  expression (\ref{TPGG_04}) is positive.
In conclusion, the aspiration-level-driven self-learning dynamics
  can afford to be less strict than the pairwise comparison process.
Namely, it requires less reward for cooperators' contribution to the common pool
  (lower levels of $r$)
  in order to promote the cooperative strategy.
The amount of cooperative strategy depends on the threshold:
  higher thresholds support cooperation,
  even for lower multiplication factors $r$
  (see Fig.~\ref{fig:TPGG}).
For fixed $r$,
  our self-learning dynamics are more likely to promote cooperation
  in a threshold public goods game,
  if the threshold for the number of cooperators needed to support the public goods
  is large enough,
  i.e., not too different from the total size of the group.
For small thresholds,
  and thus higher temptation to defect in groups with less cooperators,
  we approach the regular public goods games,
  and the conclusion may be reversed.
Under such small $m$ cases,
  imitation-driven (pairwise comparison) dynamics
  are more likely to lead to cooperation than aspiration dynamics.

\subsection{$d$-player snowdrift game}

\begin{figure}[!ht]
\begin{center}
\includegraphics[width=0.95\textwidth]{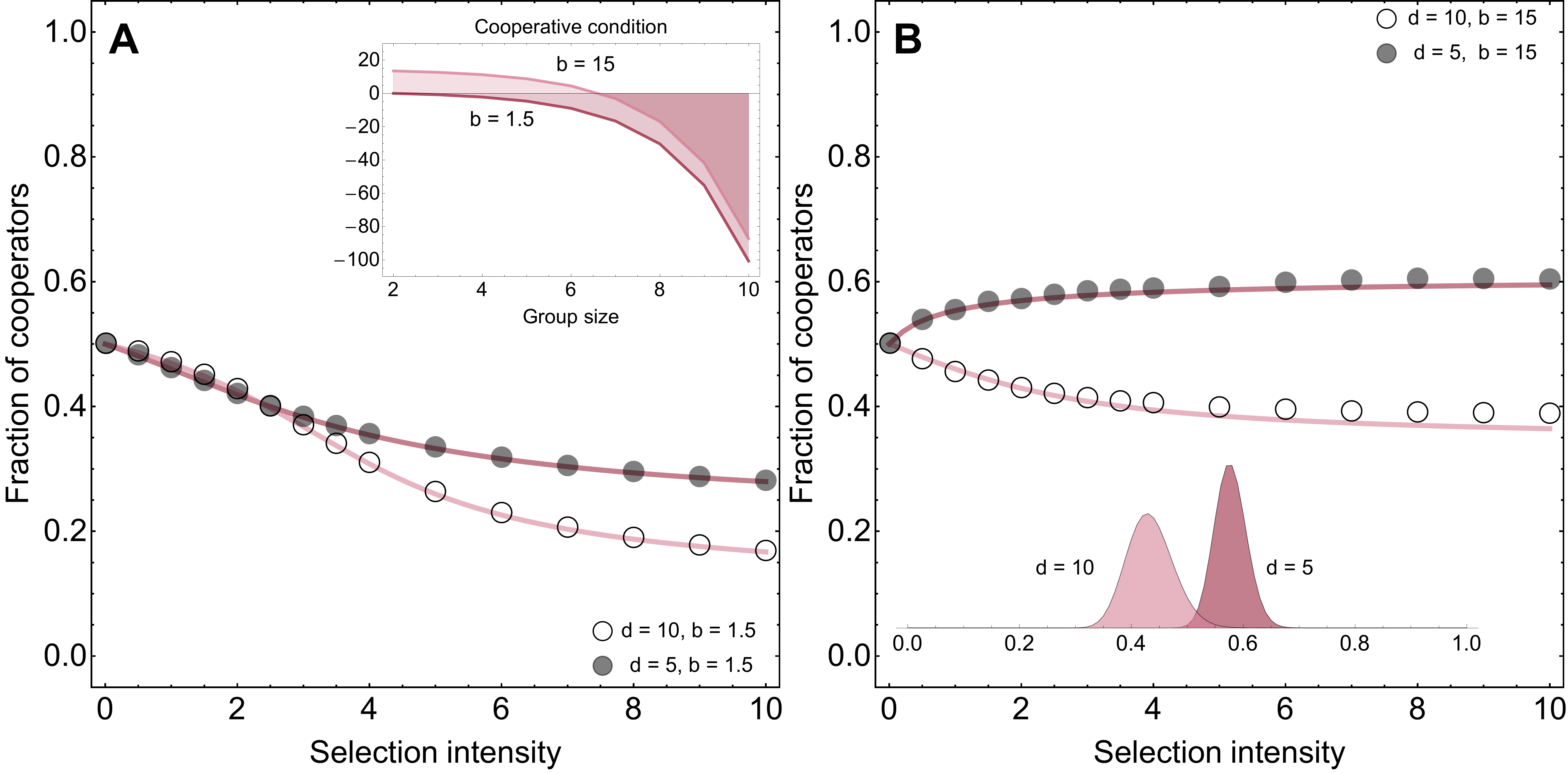}
\caption{
\small
{\bf Mean (stationary) fraction of cooperators for the $d$-player snowdrift game.}
The common parameters are
  aspiration level $\alpha=1$, population size $N=100$, and cost of cooperation $c=1$.
In both panels,
  the group sizes are $d=5$ (dark shaded), and $d=10$ (light shaded).
Panel {\bf A} shows the mean fraction of cooperators
  as a function of selection intensity for $b=1.5$.
The inset shows the cooperation condition
  as a function of group size $d$ for benefits $b=1.5,\,15.0$.
Only for high benefit and low group size,
  cooperation can be more abundant than defection.
Panel {\bf B} shows the mean fraction of cooperators
  as a function of selection intensity for $b=15.0$.
The inset shows the stationary distribution for $b=15.0$, $\omega=2$, and $d=5,\,10$.
}\label{fig:SDG}
\end{center}
\end{figure}

Evolutionary games between two strategies
  can have mixed evolutionary stable states
  \cite{Hofbauer1998Cambridge,Gokhale2010PNAS}.
Strategy $A$ can invade $B$ and $B$ can invade $A$;
  a stable coexistence of the two strategies typically evolves.
In the replicator dynamics of the snowdrift game,
  cooperators can be invaded by defectors
  as the temptation to defect is still larger
  than the reward of mutual cooperation
  \cite{doebeli:Science:2004,doebeli:EL:2005}.
In contrast to the public goods game,
  cooperation with a group of defectors now
  yields a payoff greater than exclusive defection.
The act of cooperation
  provides a benefit to all members of the group,
  and the cost of cooperation
  is equally shared among the number of cooperators \cite{Zheng2007EPL}.
Hence, the payoff matrix reads
\begin{align}
\label{SDG_Matrix}
\begin{tabular}{ l | c c c c c c c }
    & $d-1$           & $\dots$ & $k$               & $\dots$ & $1$             & $0$   \\
\hline
$A$ & $b-\frac{c}{d}$ & $\dots$ & $b-\frac{c}{k+1}$ & $\dots$ & $b-\frac{c}{2}$ & $b-c$ \\
$B$ & $b$             & $\dots$ & $b$               & $\dots$ & $b$             & $0$   \\
\end{tabular}
\end{align}
Cooperation can maintain a minimal positive payoff from the cooperative act,
  then cooperation and defection can coexist.
The snowdrift game is a social dilemma,
  as selection does not favor the social optimum of exclusive cooperation.
The level of coexistence depends on the amount of cost
  that a particular cooperator has to contribute in a certain group.
Evaluating the weak selection condition,
  (\ref{Criterion Condition}) in case of the $d$-player snowdrift game
  leads to the condition
\begin{align}
\label{SD_01}
\sum_{k=0}^{d-1}\frac{C_{d-1}^{k}}{k+1}<\frac{b}{c}
\end{align}
  in order to observe $\langle X_A\rangle>1/2$
  in aspiration dynamics under weak selection.
For imitation processes,
  on the other hand,
  we find $\sum_{k=0}^{d-1}1/(k+1)<b/c$.
Note that,
  except for $a_{0}-b_{0}=b-c>0$,
  $a_{k}-b_{k}<0$ holds for any other $k$.
Because of this,
  the different nature of these two conditions,
  given by the positive coefficients $C_{d-1}^{k}>1$ for any $d>k>0$,
  reveals that self-learning dynamics narrow down the parameter range
  for which cooperation can be favored by selection.
In the snowdrift game,
  self-learning dynamics are less likely to favor cooperation
  than pairwise comparison processes.
Larger group size $d$ hinders cooperation:
  the larger the group,
  the higher the benefit of cooperation, $b$,
  has to be in order to support cooperation (see Fig.~\ref{fig:SDG}).

\section{Summary and Conclusions}

\label{sec:Summary}
\setcounter{equation}{0}

Previous studies on self-learning mechanism
  have typically been investigated on graphs via simulations,
  which often employ stochastic aspiration-driven update rules
  \cite{WangWX2006PRE,Gao2007PA,Chen2008PRE,Liu2011EPL,Roca2011PNAS}.
Although results based on the mean field approximations are insightful
  \cite{Chen2008PRE,Szabo2007PR},
  further analytical insights have been lacking so far.
Thus it is constructive to introduce and discuss a reference case of
  stochastic aspiration-driven dynamics of self-learning in well-mixed populations.
To this end, here we introduce and discuss such an evolutionary process.
Our weak selection analysis is based on a simplified scenario
  that implements a non-adaptive self-learning process with global aspiration level.

Probabilistic evolutionary game dynamics driven by aspiration
  are inherently innovative
  and do not have absorbing boundaries
  even in the absence of mutation or random strategy exploration.
We study the equilibrium strategy distribution in a finite population
  and make a weak selection approximation for the average strategy abundance
  for any multi-player game with two strategies,
  which turns out to be independent of the level of aspiration.
This is different from the aspiration dynamics in infinitely large populations,
  where the evolutionary outcome crucially
  depends on the aspiration level \cite{Posch1999PRSB}.
Thus it highlights the intrinsic differences
  arising from finite stochastic dynamics of multi-player games between two strategies.
Based on this
  we derive a condition for one strategy to be favored over the other.
This condition then
  allows a comparison of a strategy's performance to other prominent game dynamics
  based on pairwise comparison between two strategies.

Most of the complex strategic interactions in natural populations,
  ranging from competition and cooperation in microbial communities
  to social dilemmas in humans,
  take place in groups rather than pairs.
Thus multi-player games have attracted increasing interest in different areas
  \cite{Broom1997BMB,Kurokawa2009PRSB,Pacheco2009PRSB,Souza2009JTB,Gokhale2010PNAS,Perc2013RSIF,Wu2013Games}.
The most straightforward form of multi-player games
  makes use of the generalization of the payoff matrix concept \cite{Broom1997BMB}.
Such multi-player games are more complex
  and show intrinsic difference from $2\times2$ games.
Hence, as examples here we have studied the dynamics of one of the most widely studied
  multi-player games--the linear public goods game \cite{Kurokawa2009PRSB},
  a simplified version of a threshold public goods game
  that requires a group of players to coordinate contributions to a public good
  \cite{Milinski2006PNAS,Milinski2008PNAS,santos:PNAS:2011,Du2012PRE,abouchakra:PLoSCB:2012,hilbe:PlosOne:2013},
  as well as a multi-player version of the snowdrift game \cite{Souza2009JTB}
  where coexistence is possible.
Our analytical finding
  allows a characterization of the evolutionary success
  under the stochastic aspiration-driven update rules introduced here,
  as well as a comparison to the well known results of pairwise comparison processes.
While in coordination games,
  such as the threshold public goods game,
  the self-learning dynamics support cooperation on a larger set in parameter space;
  the opposite is true for coexistence games,
  where the condition for cooperation to be more abundant becomes more strict.

It will be interesting to derive analytical results
  that either hold for any intensity of selection,
  or at least for the limiting case of strong selection
  \cite{altrock:PRE:2009,altrock:JTB:2012} in finite populations.
On the other hand,
  the update rule presented here
  does not seem to allow a proper continuous limit
  in the transition to infinitely large populations \cite{traulsen:PRL:2005},
  which might give rise to interesting rescaling requirements
  of the demographic noise in the continuous approximation
  \cite{traulsen:PRE:2012} in self-learning dynamics.

Our simple model illustrates that
  aspiration-driven self-learning dynamics in well-mixed populations alone
  may be sufficient to alter the expected strategy abundance.
On previous studies of such processes in structured populations
  \cite{WangWX2006PRE,Gao2007PA,Chen2008PRE,Liu2011EPL},
  this effect might have been overshadowed
  by the properties of the network dynamics studied {\em in silico}.
Our analytical results hold for weak selection,
  which might be a useful framework
  in the study of human interactions \cite{Traulsen2010PNAS},
  where it is still unclear to what role model
  individuals compare their payoffs
  and with what strength players update their strategies
  \cite{Traulsen2010PNAS,grujic:PLoSOne:2012,Rand2012Nature}.
Although weak selection approximations 
  are widely applied in the study of frequency dependent selection
  \cite{Lessard2007JMB,Tarnita2009JTBa,Nowak2004Nature,altrock:NJP:2009},
  it is not clear whether the successful spread of behavioral traits 
  operates in this parameter regime.
Thus, by numerical evaluation and simulations 
  we show that our weak selection predictions also hold for strong selection.
Models such as the one presented here
  may be used in attempts to predict human strategic dynamics
  \cite{nowak:JTB:2012,rand:TCS:2013}.
Such predictions,
  likely to be falsified in their simplicity \cite{grujic:PLoSOne:2010},
  are essential to our fundamental understanding of complex economic and social behavior
  and may guide statistical insights to the effective functioning of the human mind.

\section*{Acknowledgements}
This work is supported by
  the National Natural Science Foundation of China (NSFC)
  under Grants No.~61020106005 and No.~61375120.
B.W. gratefully acknowledges generous sponsorship from the Max Planck Society.
P.M.A. greatefully acknowledges support from the Deutsche Akademie der Naturforscher Leopoldina, Grant No.~LPDS 2012-12.

\newpage

\begin{appendix}

\section{Appendix}
\label{Appendix}
\setcounter{equation}{0}

In this appendix, we detail the deducing process of the criterion
  of $\langle X_A \rangle(\omega)>1/2$ for general $d$-player game.
We consider the first order approximation of stationary distribution, $\psi_{j}(\omega)$,
  and get the criterion condition (shown in Sec.~\ref{sec:Res}),
  as follows:
\begin{align}
\sum_{j=0}^{N}\frac{j}{N}\,\left[\frac{\partial}{\partial \omega}\,\psi_{j}(\omega)\right]_{\omega=0}\,\omega>0.
\label{A1}
\end{align}
Inserting Eq.~(\ref{Stationary Distribution}),
  we have
\begin{align}
\frac{\partial}{\partial \omega}\,\psi_{j}(\omega)
=
\frac
{
(\frac{\prod_{i=0}^{j-1}T_{i}^{+}}{\prod_{i=1}^{j}T_{i}^{-}})'\,
(1+\sum_{k=0}^{N-1}\frac{\prod_{i=0}^{k}T_{i}^{+}}{\prod_{i=1}^{k+1}T_{i}^{-}})
-
(\frac{\prod_{i=0}^{j-1}T_{i}^{+}}{\prod_{i=1}^{j}T_{i}^{-}})\,
(1+\sum_{k=0}^{N-1}\frac{\prod_{i=0}^{k}T_{i}^{+}}{\prod_{i=1}^{k+1}T_{i}^{-}})'
}
{
(1+\sum_{k=0}^{N-1}\frac{\prod_{i=0}^{k}T_{i}^{+}}{\prod_{i=1}^{k+1}T_{i}^{-}})^{2}
}.
\label{A2}
\end{align}
Denoting $\psi_{j}(\omega)=\psi_{N}/\psi_{D}$,
  the above equation can be simplified as
\begin{align}
\frac{\partial}{\partial \omega}\,\psi_{j}(\omega)
=\frac{\psi_{N}'\,\psi_{D}-\psi_{N}\,\psi_{D}'}{\psi_{D}^{2}},
\label{A3}
\end{align}
  where
\begin{align}
\psi_{N}'
&=(\frac{\prod_{i=0}^{j-1}T_{i}^{+}}{\prod_{i=1}^{j}T_{i}^{-}})'
=\frac{ (\prod_{i=0}^{j-1}T_{i}^{+})'\,(\prod_{i=1}^{j}T_{i}^{-})
-(\prod_{i=0}^{j-1}T_{i}^{+})\,(\prod_{i=1}^{j}T_{i}^{-})' }
{(\prod_{i=1}^{j}T_{i}^{-})^{2}}
\nonumber \\
&=\frac{
[\sum_{i=0}^{j-1}(T_{i}^{+})'\,
(\prod_{k=0,k\neq i}^{j-1} T_{k}^{+})]\,
(\prod_{i=1}^{j}T_{i}^{-})
-
(\prod_{i=0}^{j-1}T_{i}^{+})\,
[\sum_{i=1}^{j}(T_{i}^{-})'\,(\prod_{k=1,k\neq i}^{j} T_{k}^{-})]
}
{(\prod_{i=1}^{j}T_{i}^{-})^{2}},
\label{A4} \\
\psi_{D}'
&=(1+\sum_{k=0}^{N-1}\frac{\prod_{i=0}^{k}T_{i}^{+}}{\prod_{i=1}^{k+1}T_{i}^{-}})'
=\sum_{k=0}^{N-1}
\frac{
(\prod_{i=0}^{k}T_{i}^{+})'\,
(\prod_{i=1}^{k+1}T_{i}^{-})
-(\prod_{i=0}^{k}T_{i}^{+})\,
(\prod_{i=1}^{k+1}T_{i}^{-})'
}
{(\prod_{i=1}^{k+1}T_{i}^{-})^{2}}
\nonumber \\
&=\sum_{k=0}^{N-1}
\frac{
[\sum_{i=0}^{k}(T_{i}^{+})'\,
(\prod_{s=0,s\neq i}^{k}T_{s}^{+})]\,
(\prod_{i=1}^{k+1}T_{i}^{-})
-
(\prod_{i=0}^{k}T_{i}^{+})\,
[\sum_{i=1}^{k+1}(T_{i}^{-})'\,
(\prod_{s=1,s\neq i}^{k+1}T_{s}^{-})]
}
{(\prod_{i=1}^{k+1}T_{i}^{-})^{2}}.
\label{A5}
\end{align}
We have
\begin{align}
(T_{i}^{+})'
&=\frac{N-i}{N}\,
\left\{\frac{1}{1+e^{-\omega\,[\alpha-\pi_{B}(i)]}}\right\}'
=\frac{N-i}{N}\,
\frac{
\left\{e^{-\omega\,[\alpha-\pi_{B}(i)]}\right\}\,
[\alpha-\pi_{B}(i)]
}
{
\{1+e^{-\omega\,[\alpha-\pi_{B}(i)]}\}^{2}
},
\label{A6} \\
(T_{i}^{-})'
&=\frac{i}{N}\,
\left\{\frac{1}{1+e^{-\omega\,[\alpha-\pi_{A}(i)]}}\right\}'
=\frac{i}{N}\,
\frac{
\left\{e^{-\omega\,[\alpha-\pi_{A}(i)]}\right\}\,
[\alpha-\pi_{A}(i)]}
{\left\{1+e^{-\omega\,[\alpha-\pi_{A}(i)]}\right\}^{2}}.
\label{A7}
\end{align}
Since $\omega\rightarrow0$,
\begin{align}
&\left.(T_{i}^{+})'\right|_{\omega=0}=\frac{N-i}{4N}\,[\alpha-\pi_{B}(i)],
\label{A8} \\
&\left.(T_{i}^{-})'\right|_{\omega=0}=\frac{i}{4N}\,[\alpha-\pi_{A}(i)],
\label{A9} \\
&\left.(\prod_{i=0}^{j-1}T_{i}^{+})\right|_{\omega=0}
=\prod_{i=0}^{j-1}\frac{N-i}{2N}
=\frac{N!}{(N-j)!\,(2N)^{j}},
\label{A10} \\
&\left.(\prod_{i=1}^{j}T_{i}^{-})\right|_{\omega=0}
=\prod_{i=1}^{j}\frac{i}{2N}
=\frac{j!}{(2N)^{j}},
\label{A11}
\end{align}
\begin{align}
&\left.[\sum_{i=0}^{j-1}(T_{i}^{+})'\,(\prod_{k=0,k\neq i}^{j-1} T_{k}^{+})]\right|_{\omega=0}
=\frac{N!\,\sum_{i=0}^{j-1}[\alpha-\pi_{B}(i)]}{2\,(N-j)!\,(2N)^{j}},
\label{A12} \\
&\left.[\sum_{i=1}^{j}(T_{i}^{-})'\,(\prod_{k=1,k\neq i}^{j} T_{k}^{-})]\right|_{\omega=0}
=\frac{j!\,\sum_{i=1}^{j}[\alpha-\pi_{A}(i)]}{2\,(2N)^{j}}.
\label{A13}
\end{align}
Then, inserting Eqs.~(\ref{A10})--(\ref{A13}) into Eq.~(\ref{A4}),
\begin{align}
\left.\psi_{N}'\right|_{\omega=0}
&=\frac
{
\frac{N!}{2\,(N-j)!\,(2N)^{j}}\,
\left\{\sum_{i=0}^{j-1}[\alpha-\pi_{B}(i)]\right\}\,
\frac{j!}{(2N)^{j}}
-
\frac{N!}{(N-j)!\,(2N)^{j}}\,
\frac{j!}{2\,(2N)^{j}}\,
\sum_{i=1}^{j}[\alpha-\pi_{A}(i)]
}
{[\frac{j!}{(2N)^{j}}]^{2}}
\nonumber \\
&=\frac{N!}{2\,j!\,(N-j)!}\,
[\,-\sum_{i=0}^{j-1}\pi_{B}(i)+\sum_{i=1}^{j}\pi_{A}(i)\,]
\nonumber\\
&=\frac{C_{N}^{j}}{2}\,
\sum_{i=1}^{j}\left[\,\pi_{A}(i)-\pi_{B}(i-1)\,\right].
\label{A14}
\end{align}
Similarly, we can get
\begin{align}
\left.\psi_{D}'\right|_{\omega=0}
&=\sum_{k=0}^{N-1}
\{\,
\frac
{
\frac{N!}{2\,(N-k-1)!\,(2N)^{k+1}}\,
\sum_{i=0}^{k}[\alpha-\pi_{B}(i)]\,
\frac{(k+1)!}{(2N)^{k+1}}
}
{[\frac{(k+1)!}{(2N)^{k+1}}]^{2}}
\nonumber \\
&-
\frac
{
\frac{N!}{(N-k-1)!\,(2N)^{k+1}}\,
\frac{(k+1)!}{2\,(2N)^{k+1}}\,
\sum_{i=1}^{k+1}[\alpha-\pi_{A}(i)]
}
{[\frac{(k+1)!}{(2N)^{k+1}}]^{2}}
\,\}
\nonumber\\
&=\sum_{k=1}^{N}\frac{C_{N}^{k}}{2}\,
\sum_{i=1}^{k}\left[\,\pi_{A}(i)-\pi_{B}(i-1)\,\right].
\label{A15}
\end{align}
And
\begin{align}
\left.\psi_{D}\right|_{\omega=0}
&=1+\sum_{k=0}^{N-1}
\frac{\prod_{i=0}^{k} \frac{N-i}{2N} }
{\prod_{i=1}^{k+1} \frac{i}{2N}}
=1+\sum_{k=0}^{N-1}C_{N}^{k+1} =2^{N},
\label{A16} \\
\left.\psi_{N}\right|_{\omega=0}
&=\frac{\prod_{i=0}^{j-1} \frac{N-i}{2N}}
{\prod_{i=1}^{j} \frac{i}{2N}} =C_{N}^{j}.
\label{A17}
\end{align}
Therefore, inserting Eqs.~(\ref{A14})--(\ref{A17}) into Eq.~(\ref{A3}),
\begin{align}
\left[\frac{\partial}{\partial \omega}\,\psi_{j}(\omega)\right]_{\omega=0}
&=\frac {C_{N}^{j}\,
\{\sum_{i=1}^{j}[\pi_{A}(i)-\pi_{B}(i-1)]
\}\,2^{N}}
{2\,(2^{N})^{2}}
\nonumber \\
&-\frac{
C_{N}^{j}\, \sum_{k=0}^{N-1}
\{ C_{N}^{k+1}\,\sum_{i=1}^{k+1}[\pi_{A}(i)-\pi_{B}(i-1)]
\}
}
{2\,(2^{N})^{2}}.
\label{A18}
\end{align}
Combined with Eq.~(\ref{A1}), the criterion is rewritten as
\begin{align}
&\sum_{j=1}^{N}
\frac{j\,\omega}{N}\,
(\,
\frac{
C_{N}^{j}\,
\{\sum_{i=1}^{j}[\pi_{A}(i)-\pi_{B}(i-1)]
\}\,
2^{N}
}
{2\,(2^{N})^{2}}
\nonumber \\
&-\frac{
C_{N}^{j}\,
\sum_{k=0}^{N-1}
\{
C_{N}^{k+1}\,
\sum_{i=1}^{k+1}[\pi_{A}(i)-\pi_{B}(i-1)]
\}
}
{2\,(2^{N})^{2}}
\,)>0,
\label{A19}
\end{align}
  where $\pi_{A}(i)$ and $\pi_{B}(i-1)$ refer to
  Eqs.~(\ref{Payoff_C}), and (\ref{Payoff_D}).
Hence,
\begin{align}
\pi_{A}(i)-\pi_{B}(i-1)
=\sum_{k=0}^{d-1}
\frac{C_{i-1}^{k}\,C_{N-i}^{d-1-k}}{C_{N-1}^{d-1}}\,
(a_{k}-b_{k}).
\label{A20}
\end{align}
Therefore the criterion equals to
\begin{align}
&\sum_{j=1}^{N}\frac{j\,\omega\,C_{N}^{j}}{2\,N\,(2^{N})^{2}}\,
[\,\sum_{i=1}^{j} \sum_{k=0}^{d-1}\frac{C_{i-1}^{k}\,C_{N-i}^{d-1-k}}{C_{N-1}^{d-1}}\,(a_{k}-b_{k})\,2^{N}
\nonumber \\
-&\sum_{m=0}^{N-1}C_{N}^{m+1}\, \sum_{i=1}^{m+1}\sum_{k=0}^{d-1}\frac{C_{i-1}^{k}\,C_{N-i}^{d-1-k}}{C_{N-1}^{d-1}}\,(a_{k}-b_{k})\,]>0
\nonumber \\
\Longleftrightarrow
&\frac{\omega}{2\,N\,(2^{N})^{2}}\,
[\,\sum_{j=1}^{N}j\,C_{N}^{j}\,\sum_{i=1}^{j} \sum_{k=0}^{d-1}\frac{C_{i-1}^{k}\,C_{N-i}^{d-1-k}}{C_{N-1}^{d-1}}\,(a_{k}-b_{k})\,2^{N}
\nonumber \\
-&\sum_{j=1}^{N}j\,C_{N}^{j}\,\sum_{m=1}^{N}C_{N}^{m}\,\sum_{i=1}^{m}\sum_{k=0}^{d-1}\frac{C_{i-1}^{k}\,C_{N-i}^{d-1-k}}{C_{N-1}^{d-1}}\,(a_{k}-b_{k})\,]>0
\nonumber \\
\Longleftrightarrow
&\frac{\omega}{4\,N\,(2^{N})}\,
[\,\sum_{j=1}^{N}2\,j\,C_{N}^{j}\,\sum_{i=1}^{j} \sum_{k=0}^{d-1}\frac{C_{i-1}^{k}\,C_{N-i}^{d-1-k}}{C_{N-1}^{d-1}}\,(a_{k}-b_{k})\,]
\nonumber \\
-&\frac{\omega\,N\,2^{N-1}}{2\,N\,(2^{N})^{2}}\,
[\,\sum_{m=1}^{N}C_{N}^{m}\,\sum_{i=1}^{m}\sum_{k=0}^{d-1}\frac{C_{i-1}^{k}\,C_{N-i}^{d-1-k}}{C_{N-1}^{d-1}}\,(a_{k}-b_{k})\,]>0
\nonumber \\
\Longleftrightarrow
&\frac{\omega}{4\,N\,(2^{N})}\,
[\,\sum_{j=1}^{N}(2j-N)\,C_{N}^{j}\,
\sum_{i=1}^{j} \sum_{k=0}^{d-1}\frac{C_{i-1}^{k}\,C_{N-i}^{d-1-k}}{C_{N-1}^{d-1}}\,(a_{k}-b_{k})\,]>0.
\label{A21}
\end{align}
We can prove that the above inequality leads to a general criterion as follows
\begin{align}
\frac{\omega}{4\,(2^{d})}\,
\sum_{k=0}^{d-1} \left[\,C_{d-1}^{k}\,(a_{k}-b_{k})\,\right]>0.
\label{A22}
\end{align}
This is the result we want to show.
For this, we only need to demonstrate
\begin{align}
\frac{\omega}{4\,N\,(2^{N})}\,
\left[\,\sum_{j=1}^{N}(2j-N)\,C_{N}^{j}\,
\sum_{i=1}^{j} \sum_{k=0}^{d-1}\frac{C_{i-1}^{k}\,C_{N-i}^{d-1-k}}{C_{N-1}^{d-1}}\,(a_{k}-b_{k})
\,\right]
=\frac{\omega}{4\,(2^{d})}\,
\left[\,\sum_{k=0}^{d-1} C_{d-1}^{k}\,(a_{k}-b_{k})\,\right].
\label{A23}
\end{align}
This equals to
\begin{align}
\sum_{j=1}^{N}(2j-N)\,C_{N}^{j}\,\sum_{i=1}^{j} \sum_{k=0}^{d-1}\frac{C_{i-1}^{k}\,C_{N-i}^{d-1-k}}{C_{N-1}^{d-1}}\,(a_{k}-b_{k})
=2^{N-d}\,N\,\sum_{k=0}^{d-1} C_{d-1}^{k}\,(a_{k}-b_{k}).
\label{A24}
\end{align}
Since such equation should hold for any choice of ($a_{k}-b_{k}$)s,
  thus
\begin{align}
\sum_{j=1}^{N}(2j-N)\,C_{N}^{j}\,\sum_{i=1}^{j} \frac{C_{i-1}^{k}\,C_{N-i}^{d-1-k}}{C_{N-1}^{d-1}}
=2^{N-d}\,N\,C_{d-1}^{k}.
\label{A25}
\end{align}
Using the identity $\sum_{j=1}^{N}\,\sum_{i=1}^{j}=\sum_{i=1}^{N}\,\sum_{j=i}^{N}$,
  we can simplify the equivalent condition as
\begin{align}
\sum_{i=1}^{N} C_{i-1}^{k}\,C_{N-i}^{d-1-k}
\sum_{j=i}^{N}(2j-N)\,C_{N}^{j}
=2^{N-d}\,N\,C_{N-1}^{d-1}\,C_{d-1}^{k}.
\label{A26}
\end{align}
This can be easily proved through mathematical induction.

Thus, we get the criterion of $\langle X_{A}\rangle(\omega)>1/2$
  for general multi-player games as Eq.~(\ref{A22}).
We rewrite this as follows
\begin{align}
\sum_{k=0}^{d-1} C_{d-1}^{k}\,(a_{k}-b_{k})>0.
\label{A27}
\end{align}

\section{Appendix}
\label{Appendix II}
\setcounter{equation}{0}

In the following tables, we demonstrate
  how selection intensity $\omega$ and the population size $N$
  influence the evolutionary results (the average fraction of cooperators) through simulation.

\begin{table}[!h]
\caption{\small
{\bf Simulation results for Linear public goods game.}
The parameters are:
  $d$=10, $\alpha$=1, $r$=2, and $c$=1.
Under such setting,
  the criterion Eq.~(\ref{Criterion Condition}) we analytically deduced reads:
  $\sum_{k=0}^{d-1} C_{d-1}^{k}\left(a_{k}-b_{k}\right)<0$,
  which means the fraction of cooperators $\langle X_A \rangle<0.5$.
}
\begin{align}
\begin{tabular}{l|l l l l l}
   \hline
            & $\omega$=0.01 & $\omega$=0.1  & $\omega$=1    & $\omega$=5    & $\omega$=10 \\
   \hline
   $N$=50   & 0.49924   & 0.4909    & 0.43151   & 0.42745   & 0.45081 \\
   $N$=100  & 0.49885   & 0.49071   & 0.43133   & 0.42905   & 0.45519 \\
   $N$=200  & 0.4994    & 0.48993   & 0.43192   & 0.42999   & 0.45694 \\
   $N$=1000 & 0.49804   & 0.49037   & 0.43188   & 0.43076   & 0.45875 \\
   \hline
\end{tabular}
\end{align}
\end{table}

\begin{table}[!h]
\caption{\small
{\bf Simulation results for Threshold public goods game.}
The parameters are:
  $d$=10, $\alpha$=1, $c$=1, and $m$=7.
Under such setting,
  the criterion Eq.~(\ref{Criterion Condition}) reads:
  $\sum_{k=0}^{d-1} C_{d-1}^{k}\left(a_{k}-b_{k}\right)>0$,
  which means the average fraction of cooperators $\langle X_A \rangle>0.5$.
}
\begin{align}
\begin{tabular}{l|l l l l l}
   \hline
            & $\omega$=0.01 & $\omega$=0.1  & $\omega$=1    & $\omega$=5  & $\omega$=10 \\
   \hline
   $N$=50   & 0.5003    & 0.50466   & 0.56262   & 0.77205   & 0.71618 \\
   $N$=100  & 0.50079   & 0.50451   & 0.56173   & 0.7626    & 0.71218 \\
   $N$=200  & 0.50087   & 0.50467   & 0.56206   & 0.76144   & 0.71154 \\
   $N$=1000 & 0.50017   & 0.50472   & 0.56054   & 0.75942   & 0.71162 \\
   \hline
\end{tabular}
\end{align}
\end{table}

\begin{table}[!h]
\caption{\small
{\bf Simulation results for multiple Snowdrift game.}
The parameters are:
  $d$=10, $\alpha$=1, $b$=1.5, and $c$=1.
Under such setting,
  the criterion Eq.~(\ref{Criterion Condition}) reads:
  $\sum_{k=0}^{d-1} C_{d-1}^{k}\left(a_{k}-b_{k}\right)<0$,
  which means the average fraction of cooperators $\langle X_A \rangle<0.5$.
}
\begin{align}
\begin{tabular}{l|l l l l l}
   \hline
            & $\omega$=0.01 & $\omega$=0.1  & $\omega$=1    & $\omega$=5  & $\omega$=10 \\
   \hline
   $N$=50   & 0.49999   & 0.49758   & 0.469955  & 0.262065  & 0.164099 \\
   $N$=100  & 0.499545  & 0.497307  & 0.470687  & 0.261983  & 0.167228 \\
   $N$=200  & 0.499789  & 0.497168  & 0.469876  & 0.261832  & 0.168782 \\
   $N$=1000 & 0.49978   & 0.49765   & 0.46985   & 0.26296   & 0.16973 \\
   \hline
\end{tabular}
\end{align}
\end{table}

It is found that for the examples we discussed,
  namely the linear public goods game,
  the threshold collective risk dilemma
  and a multi-player snowdrift game,
  our result under weak selection can be generalized for a wide range of parameters
  (higher values of $\omega$, small and large populations).
\end{appendix}

\end{document}